\documentclass[%
preprint,
 amsmath,amssymb,
 aps,
]{revtex4-2}
\usepackage{hyperref}
\usepackage{graphicx}
\usepackage{dcolumn}
\usepackage{bm}
\usepackage{hyperref}
\usepackage{colortbl, xcolor} 
\usepackage[normalem]{ulem}

\newcommand{\micron}{\mu\textrm{m}}

\begin{document}

\preprint{APS/123-QED}

\title{Translating and evaluating single-cell Boolean network interventions in the multiscale setting}

\author{John Metzcar}
 \email{jpmetzca@iu.edu - currently jmetzcar@umn.edu}
\author{Katie Pletz}%
\author{Heber L. Rocha}
\affiliation{%
 Luddy School of Informatics, Computing, and Engineering\\
 Indiana University, \\
 Bloomington, Indiana
}%


\author{Jordan C Rozum}
\email{jrozum@binghamton.edu}
\affiliation{
 Systems Science and Industrial Engineering\\
 Binghamton University (SUNY), Binghamton, New York
}%


\date{\today}

\begin{abstract}

Intracellular networks are systems that process cellular-level information and control cell fate. They can be computationally modeled using discrete dynamical systems, in particular Boolean networks, which are implicit-time causal models of discrete binary events. Furthermore, these networks can be paired with agent-based models and embedded in computational agents to drive cellular behavior. To explore this integration, we construct a set of candidate interventions that aim to induce apoptosis in a cell-survival network of a rare leukemia using exhaustive search simulation, stable motif control, and an individual-based mean field approach (IBMFA) to drive cellular behavior. Due to inherent algorithmic limitations, these interventions are most suitable for cell-level determinations, not the more realistic multicellular setting. To address these limitations, we treat the target control solutions as putative targets for therapeutic interventions and develop a pipeline to translate them to continuous-time multicellular, agent-based models. Using Boolean network analysis, we set timings for the multicellular simulation. We set the discrete-to-continuous transitions between the Boolean network and multicellular model via thresholding and produce simple computational simulations designed to emulate situations in experimental and translational biology. These include a series of increasingly complex simulations: constant substrate gradients, global substrate pulses, and time-varying boundary value pulses. We find that interventions that perform equally well in the implicit-time single-cell setting are separable in the multiscale setting in ability to impact population growth and spatial distribution. We further analyze the results showing that the population and spatial differences arise from differences in internal dynamics (stable motif controls versus target controls) and network distance between the intervention and output nodes.  This proof of concept work demonstrates the importance of accounting for internal dynamics in multicellular simulations as well as impacts on understanding of Boolean network control. It also establishes a computational laboratory that can be used in future work to explore additional connections between cell-cell interactions and internal dynamics in the development of phenotype control strategies, presenting an opportunity to optimize experimental design prior to \textit{in vitro} and \textit{in vivo} experiments.


\end{abstract}

\maketitle


\section{Introduction}

Cell fate is determined by interactions of intracellular biochemical components among themselves and the extracellular environment.
Systems biology emphasizes these interactions, often represented using networks, as a way to understand cellular information processing.
The dynamics of regulatory networks ultimately determine the behavior of a cell in response to various stimuli and can be modeled in many ways. Qualitative causal models are especially useful when detailed time-series data are not available \cite{gjergaConvertingNetworksPredictive2020, montagudPatientspecificBooleanModels2022a, werleSystemsBiologyApproach2023}. Compared to more detailed ODE models, qualitative models are easier to build and require fewer parameters to be fitted, but are nonetheless able to describe the effects of perturbation experiments, such as gene knockout. Boolean networks (BNs), perhaps the most popular of these models, represent each gene, protein, or small molecule modeled as a time-varying binary variable that is either above (ON) or below (OFF) an unspecified threshold. They provide useful approximations to system dynamics when sigmoidal responses, which tend to be less sensitive to parameter variation, dominate behaviors. In the last 20 years, they have gained in popularity as they are relatively easy to construct and analyze \cite{schwabConceptsBooleanNetwork2020b, hemedanBooleanModellingLogicbased2022, rozumBooleanNetworksPredictive2024}.
However, additional descriptions of dynamics are sometimes required to account for effects due to a spatially heterogeneous environment or growing, interacting cell populations \cite{metzcarReviewCellBasedComputational2019a, chamseddineHybridModelingFrameworks2019}. Agent-based models (ABMs), in which autonomous agents are governed by predetermined rules, have been especially useful in describing this heterogeneity \cite{altrockMathematicsCancerIntegrating2015, metzcarReviewCellBasedComputational2019a, nortonMultiscaleAgentBasedHybrid2019}. In this work, we focus on PhysiCell~\cite{ghaffarizadehPhysiCellOpenSource2018}, in which each cell of a population is instantiated as an agent in a continuous (i.e., off-lattice) simulation domain. PhysiCell has a robust ecosystem of extensions that allow it to incorporate environmental conditions, such as diffusing substrates~\cite{ghaffarizadehBioFVMEfficientParallelized2016}, extracellular matrix \cite{metzcarSimpleFrameworkAgentbased2024},  pharmacokinetic/pharmacodynamic modeling \cite{bergmanPhysiPKPDPharmacokineticsPharmacodynamics2022}, and internal cellular dynamics \cite{ponce-de-leonPhysiBoSSSustainableIntegration2023a}. Of particular interest is the ability to integrate PhysiCell with the continuous-time BN simulator MaBoSS~\cite{stollMaBoSSEnvironmentStochastic2017} as a way to incorporate intracellular dynamics. The resulting tool, PhysiBoSS~\cite{letortPhysiBoSSMultiscaleAgentbased2019, ponce-de-leonPhysiBoSSSustainableIntegration2023a}, has been previously used to study TNF-alpha induced cell death, cell-cycle modeling \cite{rusconeBuildingMultiscaleModels2024}, and cancer therapy optimization \cite{akasiadisParallelModelExploration2022}. PhysiBoSS, however, does not currently allow for simple, high-throughput testing of intracellular interventions. Testing such interventions either systematically or using bespoke algorithms is common practice in the BN literature~\cite{zeyenTargetControlBoolean2022, suDynamicsbasedApproachTarget2020, newbyStructurebasedApproachIdentify2023, parmerDynamicalModularityAutomata2023}. The link between these agent-level interventions and population-level effects is not well-studied.

In this work we provide a proof of concept solution to address some of these limitations. We determine apoptosis inducing target control solutions for the T-cell large granular lymphocytic (T-LGL) leukemia model of Zhang et al.\cite{zhangNetworkModelSurvival2008} and treat them as putative targets for therapeutic interventions. We then developed a pipeline to adapt these cell-level interventions to multicellular and multiscale simulation and produced simple computational models designed to emulate situations in experimental and translational biology.
We found that interventions equal in the single-cell setting are separable in the multiscale setting in terms of ability to limit growth of a population of simulated cells. We further analyze the results showing that the population and spatial differences arise from differences in internal dynamics (stable motif controls versus target controls) and network distance between the intervention and output nodes
Finally, we observed that these properties (network distance and nature of the control) were also predictive of efficacy in terms of infiltration of cells into the zone of lethality of an intervention. Putting this all together produces a proof of concept computational laboratory to both develop phenotype control strategies and evaluate their robustness in the multicellular setting.

\section{Methods}


\subsection{Boolean networks and their control}
\label{sec:bn-killers}
BNs are discrete dynamical models in which a network of nodes interact via directed edges~\cite{schwabConceptsBooleanNetwork2020b}. Each node $i$ is assigned a binary state variable $x_i(t)$ and a Boolean regulatory function $f_i(\boldsymbol{x})$ that depends on the states of the regulator nodes of $i$ (i.e., nodes with edges to node $i$). At each discrete time step $t$, some subset of nodes is selected to be updated, and each selected node $i$ is updated so that $x_i(t+1)=f_i(\boldsymbol{x}(t))$. The manner in which nodes are chosen for update is called the \emph{update scheme}. Two update schemes are especially common: synchronous update, in which all nodes are updated at each time step; and general asynchronous update, in which one randomly selected node is updated at each time step.

BNs of cell processes typically model cell phenotypes as the network's dynamical attractors~\cite{rozumBooleanNetworksPredictive2024}. In the context of BNs, these are typically defined as the minimal sets of states from which no escape is possible. Sometimes, however, the application is concerned with the value of a single node (or small set of nodes) in the network, the output nodes, which distinguishes biologically important attractors. Driving the system into a desired behavior can thus be approached in two different ways: attractor control or target control. In attractor control, the aim is to override the states of nodes such that the system eventually converges into one of its uncontrolled attractors. In target control, the aim is to override nodes to drive the system into a state (or small set of states) in which the output node takes on the desired value.
One successful approach to attractor control in BNs is stable motif analysis, in which key regulatory feedback loops are identified and overridden. The software package pystablemotifs~\cite{rozumParityTimeReversal2021a, rozumPystablemotifsPythonLibrary2022} implements such a procedure to identify node values that, when fixed, result in convergence to a specified attractor. The method of pystablemotifs is exact, and produces interventions that are guaranteed to result in the specified outcome.

In contrast, mean-field approaches to BN analysis are approximate. In the individual mean-field based approach (IBMFA) from Parmer et al. (2022), the BN is recast as a simpler dynamic system that uses only local interactions for node updating \cite{parmerInfluenceMaximizationBoolean2022}. Each node is initially assumed to have uniform propensity (a continuous value representing the likelihood of the node being on in the original system) of being off or on (unless otherwise specified). At each time step, the on/off propensity for each node is updated using only local inputs. This process repeats until the values reach a steady-state.

Recently, interest has grown in edgetic control, in which node values are not overridden directly, but instead the value to a single regulatory function is altered (a single outgoing edge is perturbed) while inputs to other regulatory functions continue to use the unmodified value~\cite{campbellEdgeticPerturbationsEliminate2019}. Such interventions model blocking binding sites, for example. 

\subsection{T-LGL leukemia network}

\begin{figure}
    \centering
    \includegraphics[width=1.0\linewidth]{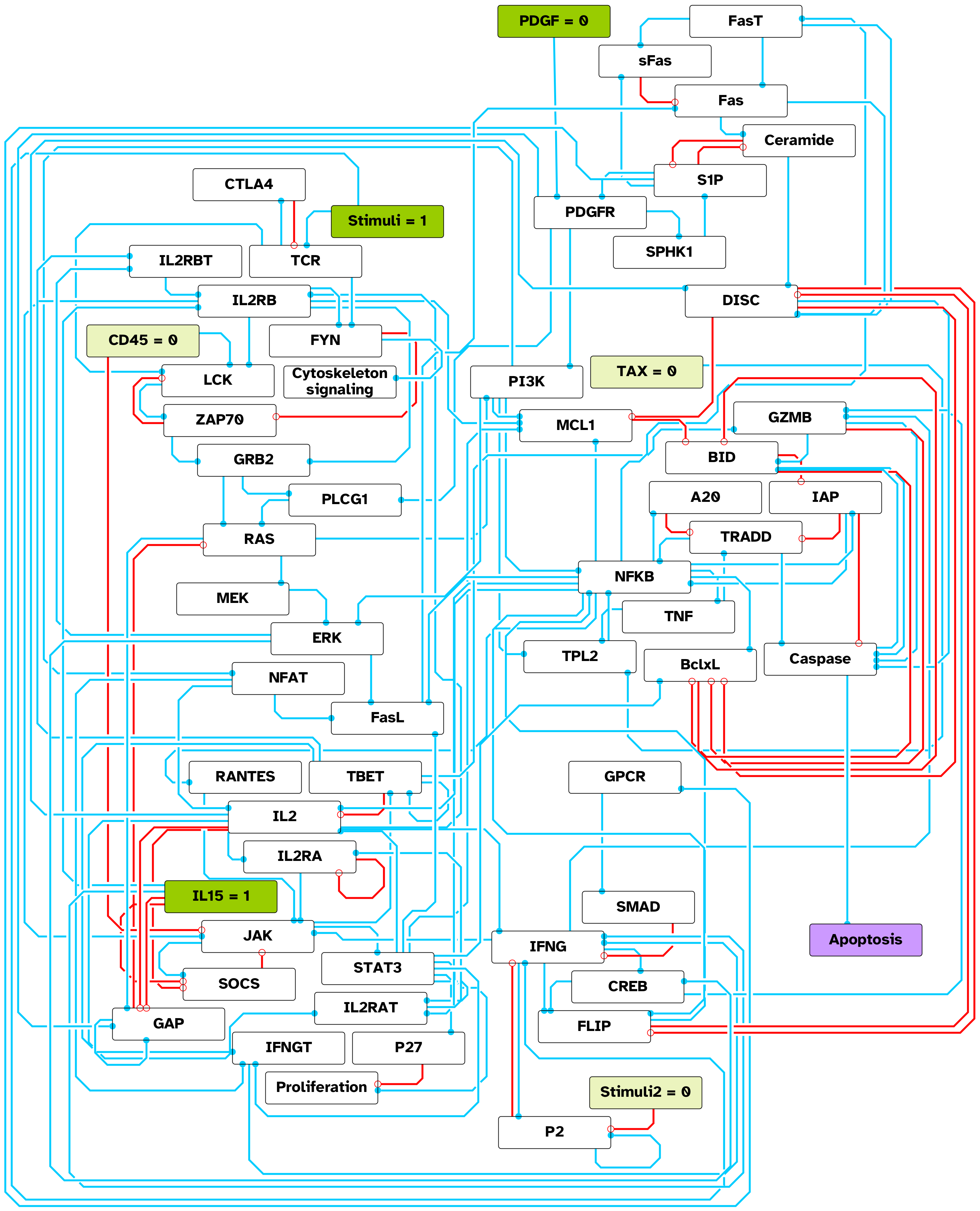}
    \caption{Wiring diagram of the TLGL leukemia network of Zhang et al \cite{zhangNetworkModelSurvival2008}. Input nodes set to produce context in which both Apoptosis ON and OFF are attainable steady states. Input nodes colored by state (light green: OFF, green: ON). Output node Apoptosis labeled in purple.}
    \label{fig:TLGL_wiring_diagram}
\end{figure}

The T-LGL leukemia network (shown in Figure \ref{fig:TLGL_wiring_diagram}) describes the dynamics of cell survival in the context of a rare leukemia of the T-cells \cite{zhangNetworkModelSurvival2008}. In the non-disease setting, the cell of origin naturally commits apoptosis. In the disease state, cells remain alive and proliferative in a non-homeostatic fashion. The model captures both non-homeostatic cell survival as well as cell death (apoptosis). This enables the study of the balance of factors that can type the cell into apoptosis. We work with network inputs set to produce a bi-stable switch in which both cell viability and apoptosis phenotypes are possible as determined by the output node Apoptosis. We take Apoptosis-ON to indicate  cellular conditions strongly favoring apoptosis and Apoptosis-OFF to represent continued viability. 

\subsection{Methods and results for candidate interventions}

We apply three control methods to the T-LGL model: the stable motif control algorithm of~\cite{zanudoCellFateReprogramming2015} implemented in pystablemotifs, target control estimated by brute force using IBMFA~\cite{parmerInfluenceMaximizationBoolean2022}, and edgetic control identified by exhaustive search using the GPU accelerated BN simulator cubewalkers~\cite{parkModelsCellProcesses2023}. The results of each algorithm were handled separately. All stable motif controls were used and edgetic controls that resulted in the highest mean asymptotic value of apoptosis in cubewalkers simulations. In the case of the IBMFA simulations, many possible overrides were returned. To reduce the number of candidates, we consider only those single-node overrides that lead to maximum apoptosis and which lead to minimal apoptosis when inverted. In all, we compile 28 unique interventions to evaluate.

\subsection{Boolean networks in multicellular simulations}


PhysiBoSS \cite{ponce-de-leonPhysiBoSSSustainableIntegration2023a, letortPhysiBoSSMultiscaleAgentbased2019} is a combined agent-based and BN model simulator. Its components are PhysiCell \cite{ghaffarizadehPhysiCellOpenSource2018}, a lattice-free agent-based modeling simulator and MaBoSS \cite{stollMaBoSSEnvironmentStochastic2017}, a stochastic BN simulator that produces continuous-time-like simulation trajectories. 


To evaluate the interventions, we make a set of BN variants. We add one new input node per intervention (\textit{e. g.} if the intervention is a combination of three nodes, we add three input nodes). The new input node (intervention node) overrides the logic of its respective node or edge such that when the input node is active, the target node is ON (a promoting intervention) or OFF (an inhibitory intervention), but when the intervention node is off, the dynamics of the target node update as prescribed in the original BN. Edge interventions are implemented similarly. When the intervention node is on, the target node's dynamics update as if the intervention edge is ON or OFF depending on the intervention type. When the intervention node is OFF, the target node updates  without alteration to dynamics. 


PhysiBoSS inherently uses a continuous, global simulation time. However, the T-LGL leukemia network does not have timings associated with node updates, as is common with BNs \cite{helikarCellCollectiveOpen2012}. To integrate this network into PhysiBoSS we simulated each intervention in MaBoSS using asynchronous BN updates. The simulations all reached apoptosis in approximately 24 simulated hours or less. Thus, as in previous modeling efforts \cite{ponce-de-leonPhysiBoSSSustainableIntegration2023a}, we match the duration of the transition from the initial survival network configuration to an apoptotic output state to 24 hours within PhysiBoSS accomplished by setting the intracellular scaling factor to 60. With this setting, a population of cells dies out in 24 hours or less, regardless of exact intervention, if constantly exposed to a pro-apoptotic substrate. 


Within PhysiBoSS, one has to specify how cell and/or environmental conditions impact the BN, as well as how the BN impacts cell behaviors or fate. Furthermore, given the discrete nature of BNs, this is often a continuous-to-discrete transition and correspondingly a discrete-to-continuous transition from the network to the cells. For the cell/environment to BN direction, we use a threshold model. If the value of the pro-apoptotic substrate is above a threshold (0.5), then the intervention is considered active and the intervention input node is set to ON. We then let the network update, holding the intervention node to ON until the substrate drops below the threshold. For our BN to cell coupling, since our example network is a survival network, we model it as controlling the apoptotic rate. When the output node Apoptosis is on, the apoptotic rate increases by a prespecified amount and returns to its regular value when Apoptosis is off. 

We use three simulations to explore the selected target control measures: \textit{constant gradient}, \textit{global pulses}, and \textit{time-varying boundary condition}. In all scenarios, cells proliferate with an average of a five day doubling time (proliferation rate of 0.00009625 min$^{-1}$). Additionally, proliferation is modeled as declining naturally as cells sense pressure. This is modeled with cell pressure as the input to a Hill function with a half-maximum of 0.75 and Hill exponent of four decreasing entry into the cell cycle, implemented through PhysiCell rules \cite{johnsonDigitizeYourBiology2023a}. The agents are initially phenotypically uniform within each experiment, starting with an individual copy of the simulation specific variant of the T-LGL leukemia cell survival network set randomly in one of two survival attractors. 

In the \textit{constant gradient} scenario, pro-apoptotic substrates are introduced via a constant Dirichlet boundary condition of 1.0 on the left-hand side of the rectangular simulation domain (400 by 800 $\micron$). The domain begins initially filled with 1344 agents in hexagonal packing configuration. After the initial transient, this produces a constant substrate gradient across the domain with substrate concentration dropping below the activation threshold at approximately the center of the domain (x=0) with a diffusion coefficient of 1025 $\frac{\micron^2}{min}$ and decay of 0.00475 min$^{-1}$. The apoptotic rate is 0.01 min$^{-1}$ when Apoptosis is on and zero otherwise. Each intervention was run once. The simulations were run to 96 hours (four days) with an approximate wall time of two minutes, splitting the 28 simulations over six compute nodes in a high-performance computing environment. 

In the \textit{global pulses} scenario, pro-apoptotic substrates are introduced uniformly with concentration of 1.0 across an 800 by 800 $\micron$ domain every 24 hours (including at time zero). The simulation begins with 1500 cells seeded randomly through the domain, again with the embedded BN controlling the apoptotic rate (set to 0.001 min$^{-1}$ when Apoptosis is on, and zero otherwise). Three decay rates were tested: 0.0116, 0.00192, 0.000963 min$^{-1}$. These rates produce effective pulse widths of one, six, and 12 hours respectively in which the apoptotic substrate is above the activation threshold. Each intervention was run at each decay rate in triplicate for a total of 252 simulations. The simulations were run to 132 hours (5.5 days) with an approximate wall time of 25 minutes in a high-performance computing environment, splitting the simulations over six compute nodes.  

In the \textit{time-varying boundary condition} scenario, pro-apoptotic substrates are introduced via a time-varying Dirichlet boundary condition on left-hand side of the rectangular simulation domain (400 by 800 $\micron$). Employing the PhysiCell addon PhysiPKPD\cite{bergmanPhysiPKPDPharmacokineticsPharmacodynamics2022}, the boundary value decays exponentially and resets every 24 simulated hours, producing waves of pro-apoptotic substrate entering the domain with the zone of lethality extending up to approximately the center of the domain (x=0). Substrate diffusion and decay are the same as in the \textit{constant gradient} scenario. Three rates of boundary value decay were tested: 0.0116, 0.00192, 0.000963 min$^{-1}$. These correspond to time above threshold (at the boundary) of one, six, and 12 hours. The domain begins initially filled with 1344 agents in a hexagonal packing configuration. The apoptotic rate is again controlled by the BN with rate set to 0.01 min$^{-1}$ when Apoptosis is ON, zero otherwise. Each intervention and rate of boundary value decay combination was run once for a total of 84 simulations. The simulations were run to 168 hours (seven days) with an approximate wall time of 25 minutes in a high-performance computing environment, splitting the simulations over six compute nodes.  

\begin{figure}
    \centering
    \includegraphics[width=1.0\linewidth]{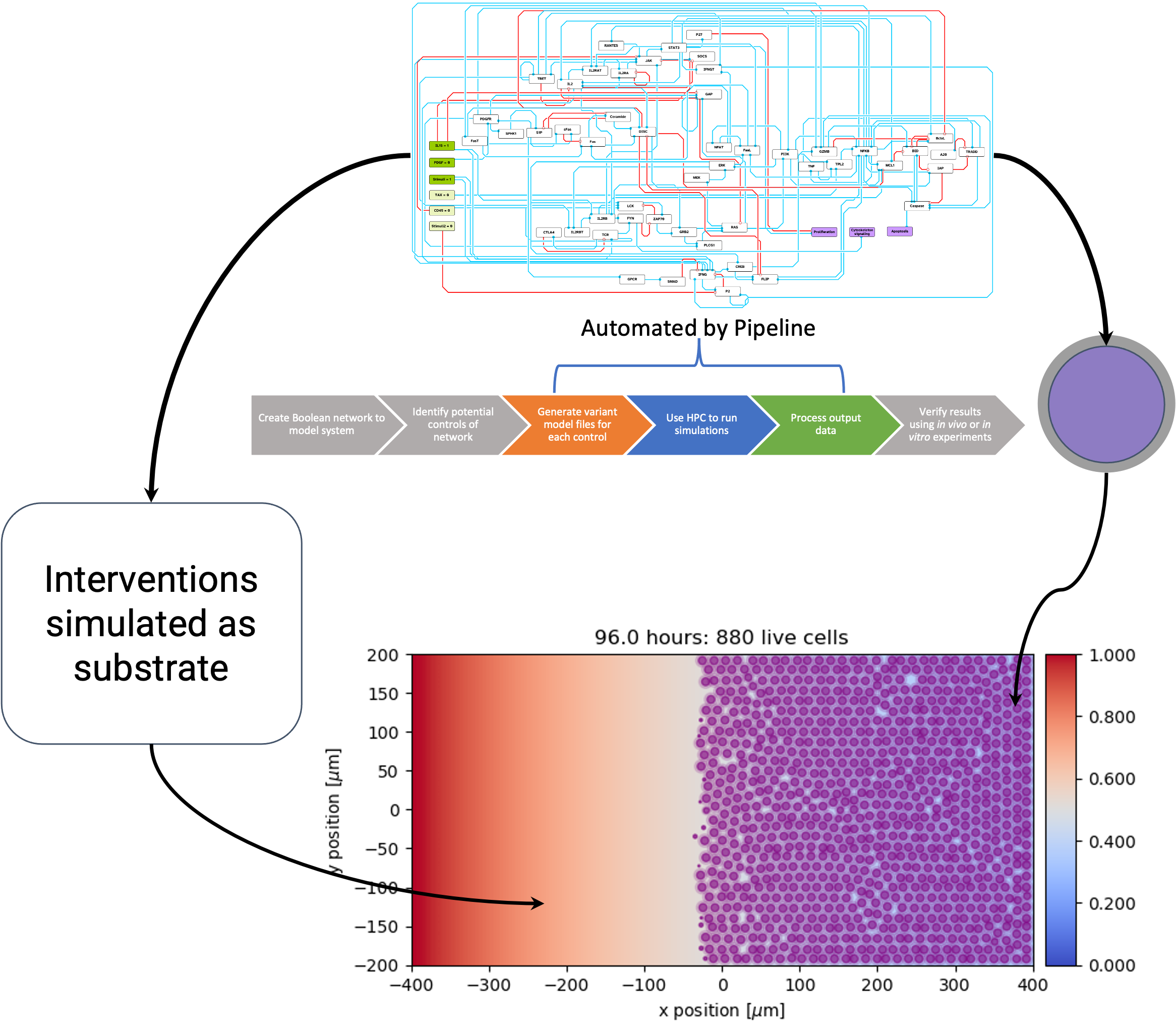}
    \caption{Schematic of the PhysiLab system and pipeline. Using target control analysis of a BN to generate interventions, we produce a set of modified BN network and PhysiBoSS simulation files. An automated pipeline produces the multiscale simulation files for running agent-based model simulations with BN controlled internal agent dynamics (via \cite{ponce-de-leonPhysiBoSSSustainableIntegration2023a}) on high-performance computing in high-throughput as well as processes simulation output.}
    \label{fig:physilab_schematic}
\end{figure}

\subsection{Code and data availability}

Our code base is available at \href{https://github.com/johnmetzcar/PhysiLab}{https://github.com/johnmetzcar/PhysiLab}. It has been tested at the basic simulation level on MacOS Monterey and Sonoma on the Intel64 chip set, Ubuntu 22.04, and Windows Server 2022. High-through put simulations were run on Linux in an HPC environment. Simulation data is available upon reasonable request. 

\section{Results}
\subsection{Overview of simulation outputs}

Interventions that produce identical long-term behavior in the single-cell BN simulation context can give rise to different long-term behaviors in the multicellular simulation context. We tested this by exposing phenotypically homogeneous cell-agents to time and spatially varying apoptosis inducing conditions. We record the location, size and other cell-variables as well as substance concentration values at a 12 minute sampling rate. This enables analysis of the impact of the different experiments on cell-fate. 

\begin{figure}
    \centering
    \includegraphics[width=0.8\linewidth]{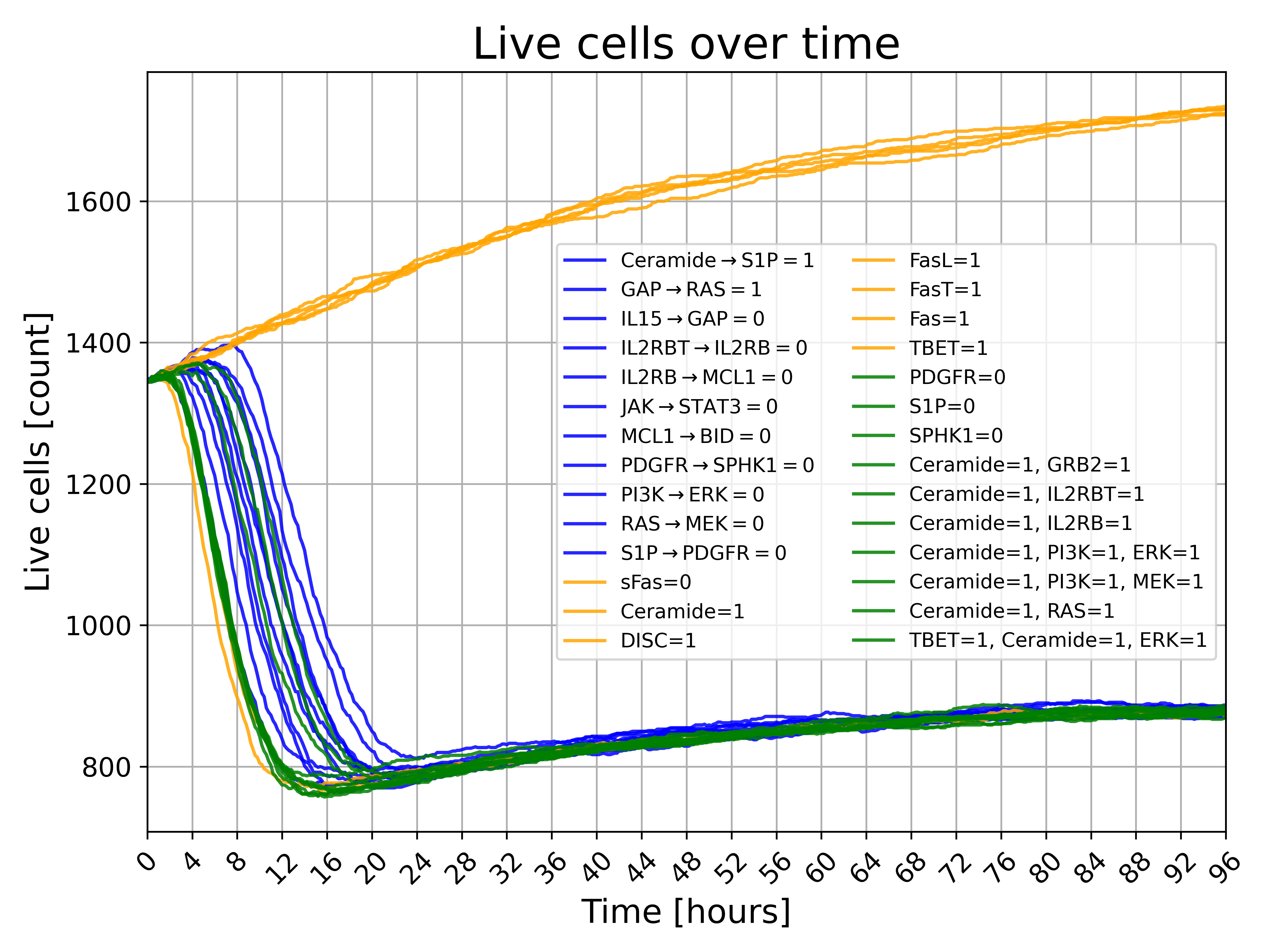}
    \caption{Cell count over time in the \emph{constant gradient} simulation scenario. Time courses are colored according to the method used to obtain the putative apoptotic target: brute force edgetic control (blue), brute force mean-field approximation control (gold), or stable motif control (green).}
    \label{fig:constant-spatial}
\end{figure}

In the first simulation scenario (\emph{constant gradient}), we expose agents to a constant gradient of the apoptotic interventions inferred using the three methods described in section~\ref{sec:bn-killers}. The population time series are plotted in Figure~\ref{fig:constant-spatial}. Notably, several of the interventions inferred using the IBMFA approach fail to suppress the population. In these cases, the interventions are only effective in the mean-field approximation and not in the true dynamics. All interventions obtained using cubewalkers or pystablemotifs result in a population reduction. Among the effective interventions, differences in time required to initiate apoptosis are readily apparent and discussed in greater detail in Section \ref{sec:network_distance_results}. The maximum rates of population decline and the asymptotic population levels, however, are similar for all effective interventions.

\begin{figure}
    \centering
    \includegraphics[width=0.8\linewidth]{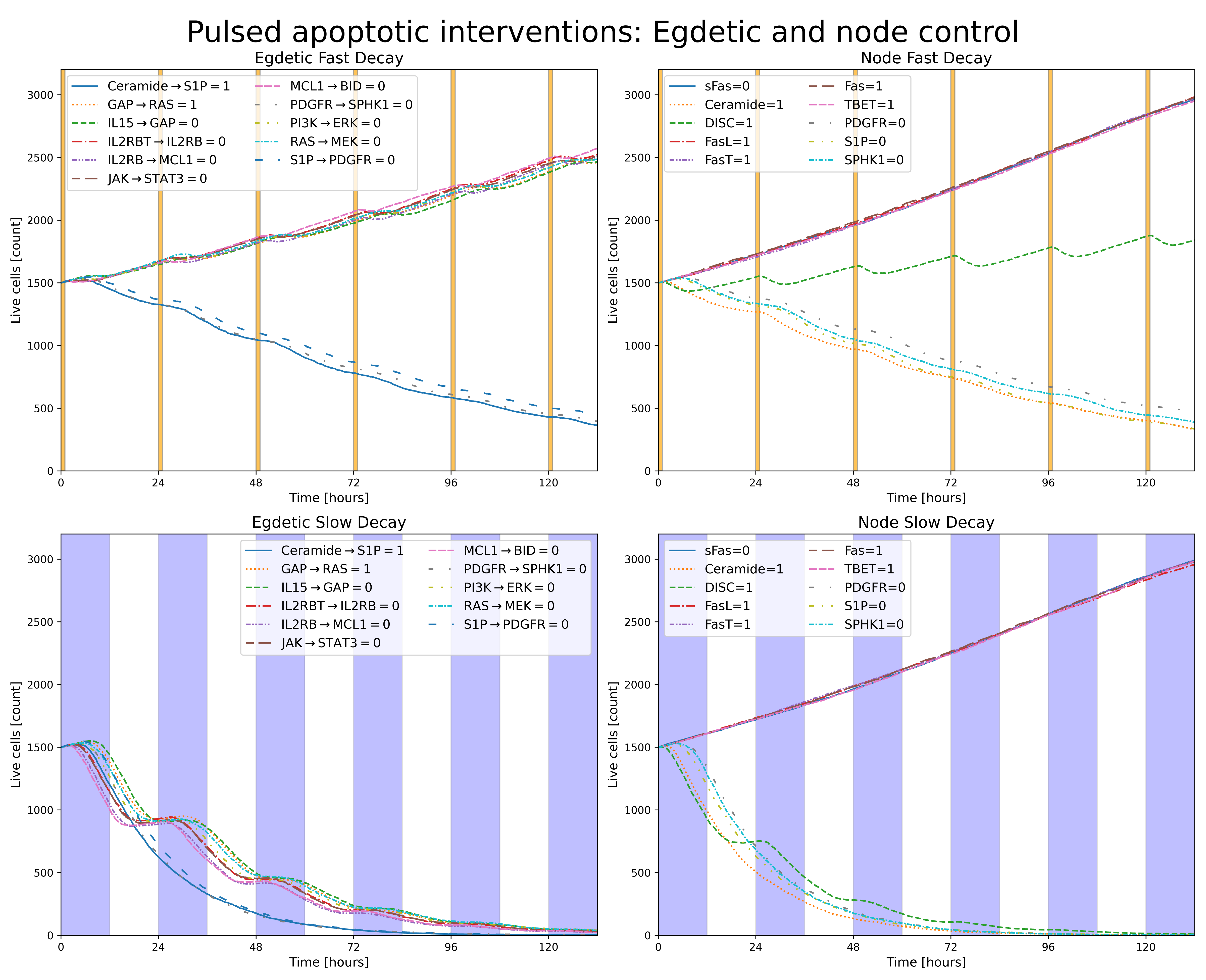}
    \caption{Population over time in four sets of experiments in the \emph{global pulses} scenario. Time courses are colored and styled according to the target of the apoptotic agent introduced. The left two panels depict interventions that target interactions, and the right two panels depict interventions that target regulatory nodes directly. In the top two panels, a fast decay for the global pulse is used, and a slow decay is used in the bottom two panels. The shaded (not white) regions indicate the periods during which the concentration of the apoptotic substrate exceeds the threshold required for effect on the internal dynamics of the cellular agents.}
    \label{fig:pulsed}
\end{figure}

\begin{figure}[htpb]
    \centering
    \includegraphics[width=0.95\linewidth]{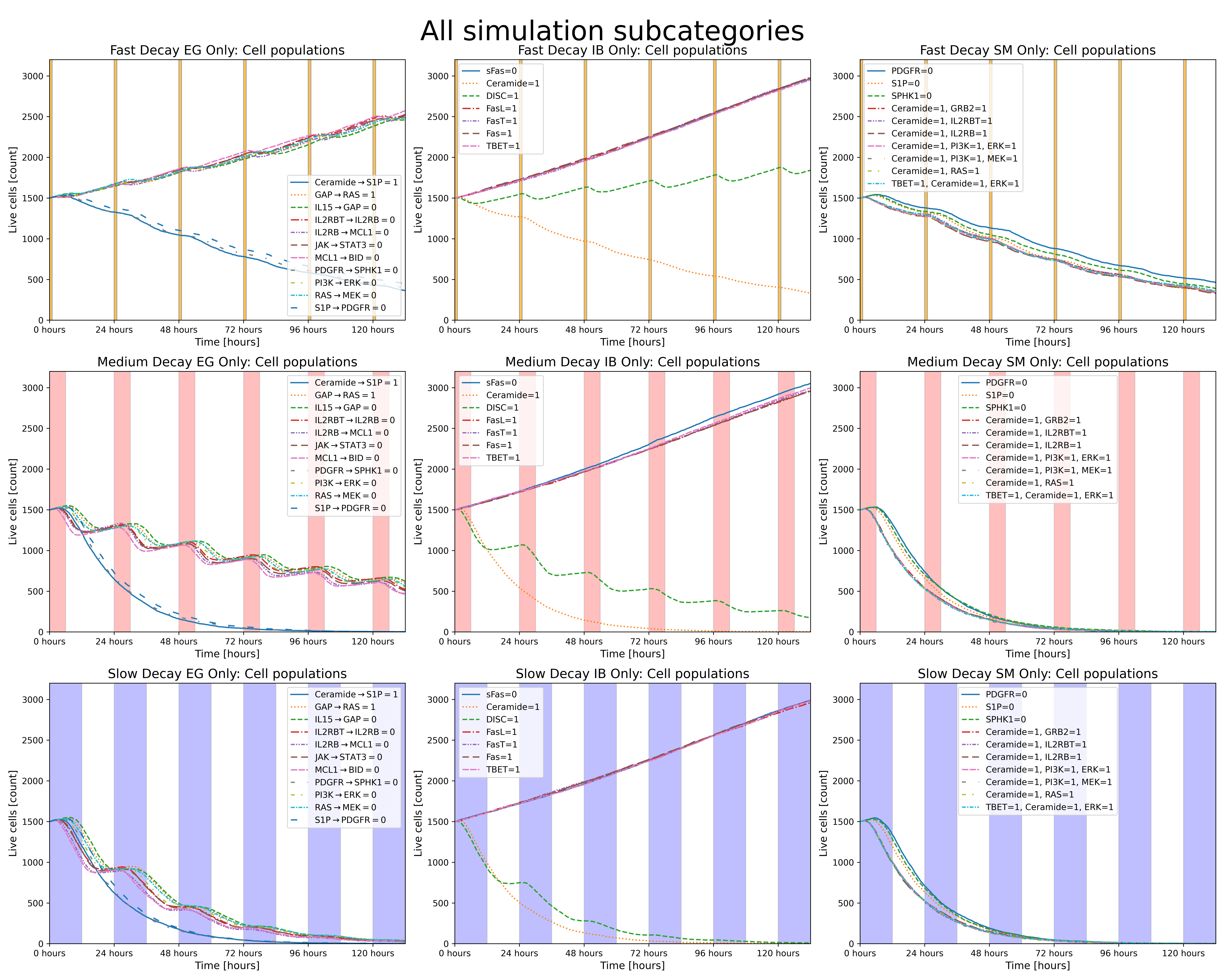}
    \caption{Population over time in all experiments in the \textit{global pulses} scenario. Time courses are colored and styled according to the target of the apoptotic agent introduced. The left column depicts interventions that target interactions (edgetic control), the middle column shows interventions determined from the results of the IBMFA method, and the right column shows interventions determined by the stable motif method. The shaded (not white) regions indicate the periods during which the concentration of the apoptotic substrate exceeds the threshold required for its effect to be registered in the internal dynamics of the cellular agents.} 
    \label{fig:all_time_simulations}
\end{figure}

In the second simulation scenario (\emph{global pulses}), we vary a spatially uniform concentration of apoptotic substrates in time using one of three decay rates and plot the population time series (see Figure~\ref{fig:pulsed}). The fast decay corresponds to a duration of lethality of one hour, the medium decay to a duration of six hours, and the slow decay to a duration of 12 hours. In the slow decay experiments, all interventions that were successful in the single agent testing (MaBoSS testing) are effective in driving the population level toward zero. In the fast decay experiments, most interventions are not successful and the population increases. In section~\ref{sec:motif-effects}, we will discuss the features of the interventions that lead to these different behaviors. See the SM Figure~\ref{fig:all_time_simulations} for full simulation results. 

\begin{figure}
    \centering
    \includegraphics[width=0.75\linewidth]{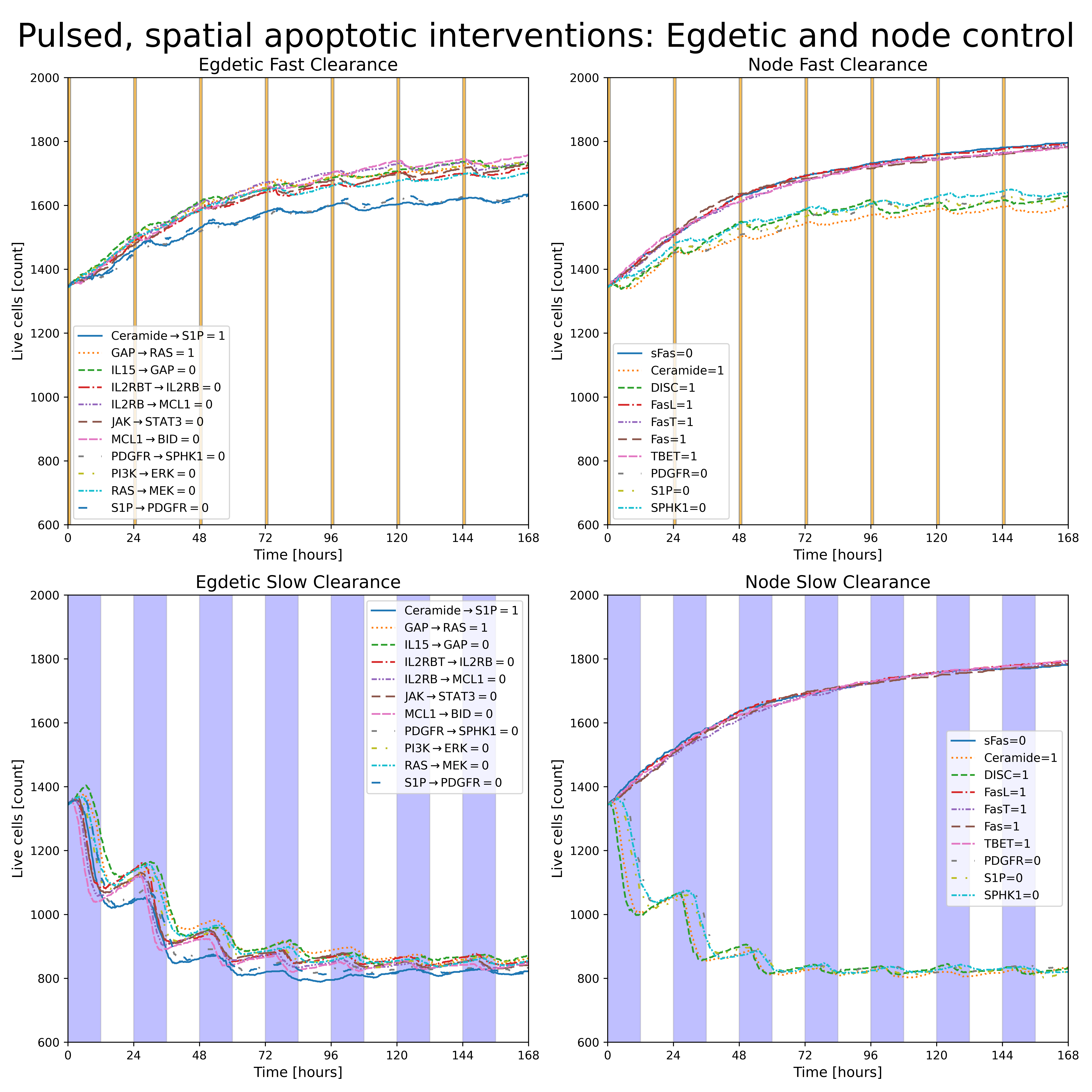}
    \caption{Population over time in four sets of experiments in the \emph{time-varying boundary condition} scenario. Time courses are colored and styled according to the target of the apoptotic agent introduced. The left two panels depict interventions that target interactions, and the right two panels depict interventions that target regulatory nodes directly. In the top two panels, a fast boundary value decay for the diffusing substrate is used, and a slow boundary value decay is used in the bottom two panels. The shaded (not white) regions indicate the periods during which the boundary of the simulation domain has an apoptotic substrate concentration above the threshold required for its effect to be registered in the internal dynamics of the cellular agents.}
    \label{fig:varying-bcs}
\end{figure}

\begin{figure}
    \centering
    \includegraphics[width=0.9\linewidth]{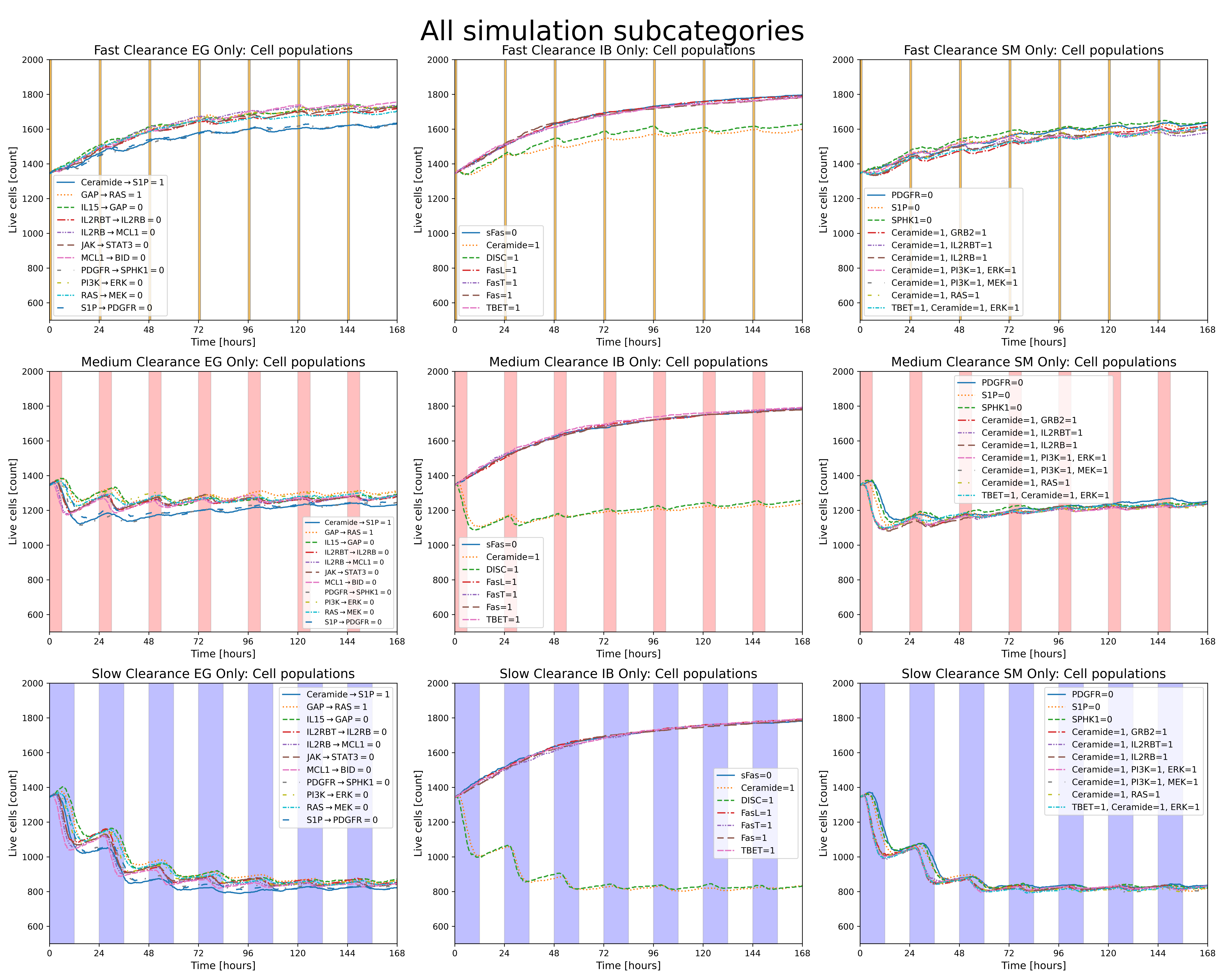}
    \caption{Population over time in all experiments in the \emph{time-varying boundary condition} scenario. Time courses are colored and styled according to the target of the apoptotic agent introduced. The left column depicts interventions that target interactions (edgetic control), the middle column shows interventions determined from the results of the IBMFA method, and the right column shows interventions determined by the stable motif method. In the top row, a fast clearance for the diffusing substrate is used, medium for the middle row, and a slow clearance is used in the bottom row. The shaded (not white) regions indicate the periods during which the boundary of the simulation domain has an apoptotic substrate concentration above the threshold required for its effect to be registered in the internal dynamics of the cellular agents.}
    \label{fig:all-varying-boundary}
\end{figure}

In the third scenario (\emph{time-varying boundary condition}), rather than globally varying a uniform concentration of an apoptotic substrate, we vary the left-hand boundary condition value in time and allow the apoptotic substrate to diffuse into the domain. The diffusion rates and boundary condition values are chosen so that half of the simulation domain experiences substrate values above the intervention activation threshold, while the other half always remains below the threshold. The population time series for different boundary value decay rates are shown in Figure~\ref{fig:varying-bcs}. The overall behavior of effective interventions is broadly similar in the experiments for this scenario, though there is greater separation in the final population levels than in the other simulation scenarios. See SM Figure \ref{fig:all-varying-boundary} for complete simulation results. 

\subsection{Measures of intervention efficacy agree across scenarios}



\begin{figure}
    \centering
    \includegraphics[width=1.0\linewidth]{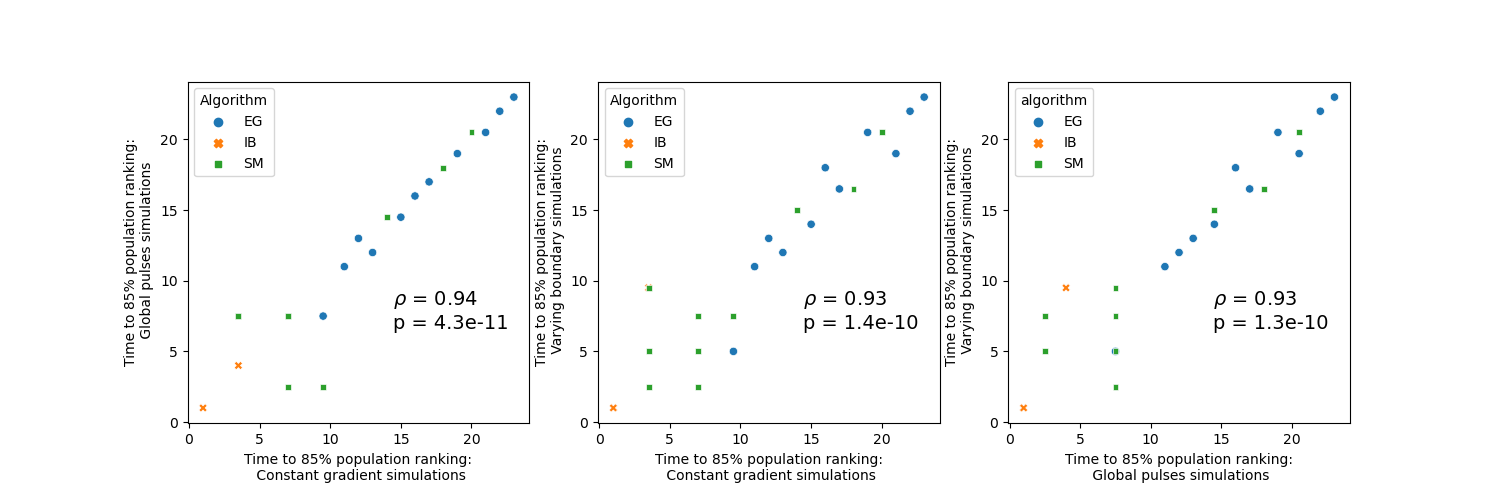}\\
    \includegraphics[width=1.0\linewidth]{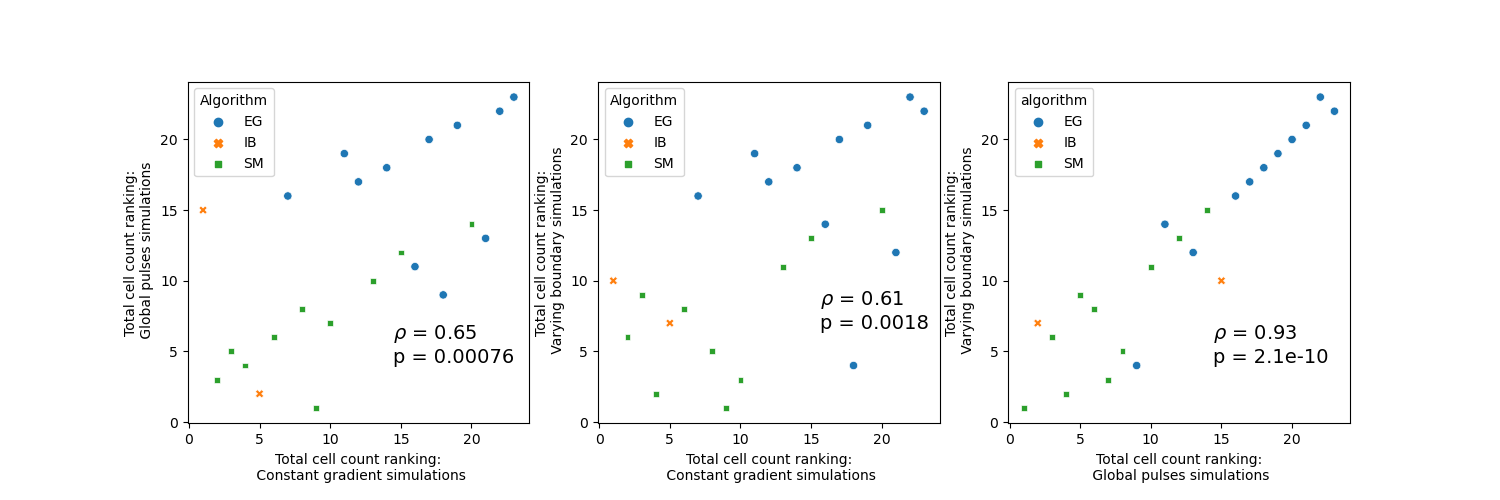}
    \caption{Cross scenario efficacy comparisons: Cross scenario efficacy ranking correlations. From left to right, \emph{constant gradient} vs. \emph{global pulses}, \emph{constant gradient} vs. \emph{time-varying boundary condition}, \emph{global pulses} vs. \emph{time-varying boundary condition}. Markers indicate intervention selection method. We used the Spearman's rank correlation test to assess the relationships. The correlation coefficient $\rho$ and associated p value are reported in each panel. \textit{Top row}: Time required to reduce population to 85\% of initial level. \textit{Bottom row}: Integrated population over simulation period. }
    \label{fig:rank-comparison}
\end{figure}

We consider two ways to measure efficacy of an intervention in each scenario: the time required to reduce the population to 85\% of its initial level, and the integrated population over the simulation period. In both cases, a lower value corresponds to a more efficacious intervention. Each intervention thus receives two rankings in each set of experiments. In Figure~\ref{fig:rank-comparison}, we show how these rankings compare across simulation scenarios. Generally, there is good agreement, though the integrated population rankings for the first scenario (\emph{constant gradient}) do not agree well with those of the other two scenarios (second row of Figure~\ref{fig:rank-comparison}). This is likely because the integrated population level in the \emph{constant gradient} scenario is primarily driven by the time required to initiate population decline, whereas in the other two scenarios, population recovery after decay of the pro-apoptotic substrate plays a role (as we discuss further in section~\ref{sec:motif-effects}).

\subsection{Stable motifs and target controls}
\label{sec:motif-effects}

\begin{figure}
    \centering
    \includegraphics[width=0.5\linewidth]{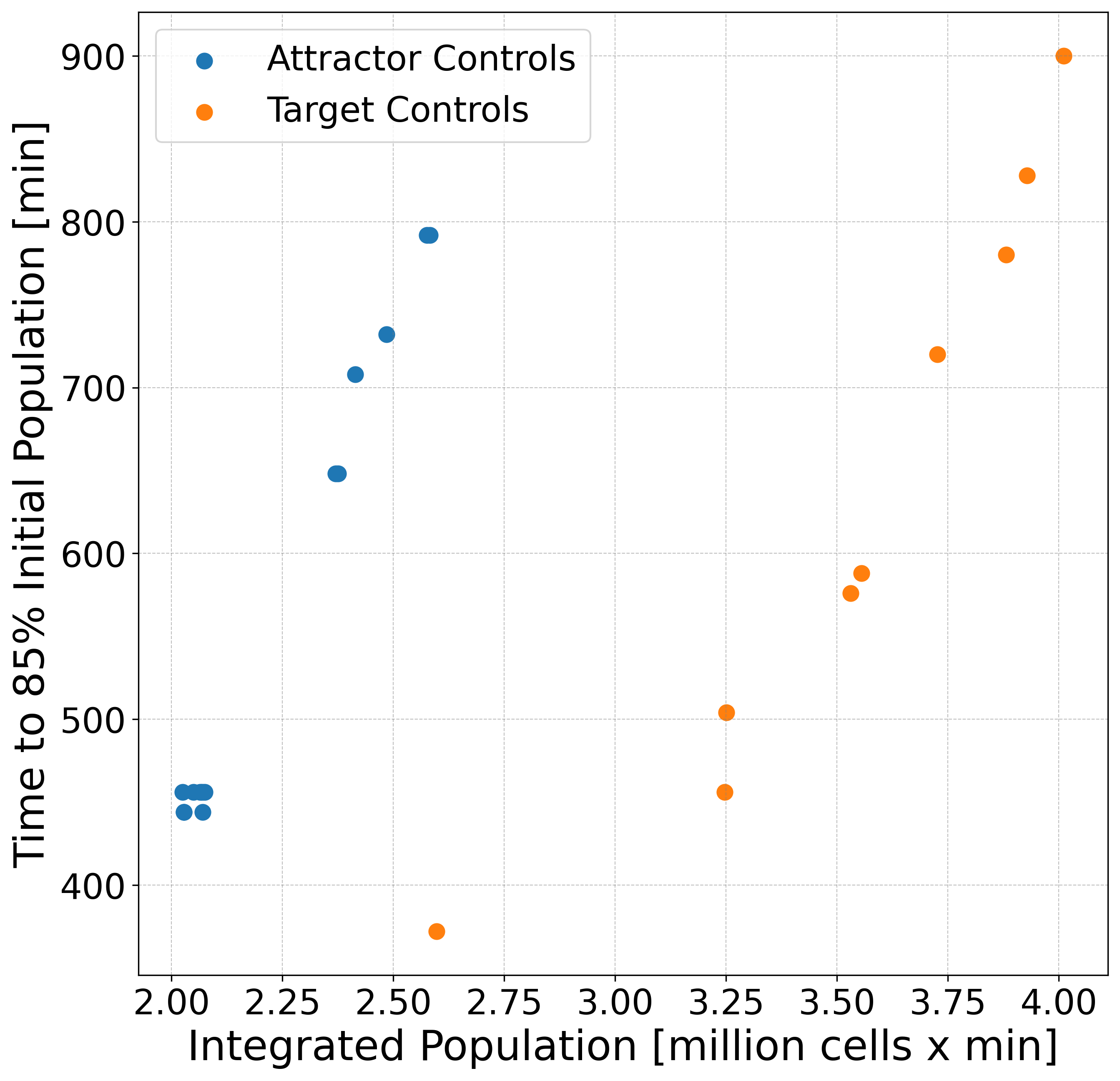}
    \caption{Relationship between initial and overall lethality in the \emph{global pulses} scenario (slow decay). Each point corresponds to the population response to an intervention averaged over three stochastic replicates. The horizontal position of each point is computed as the time integral of average cell population over the entire simulation period for an intervention. The vertical position is computed as the time at which the average population first reaches 85\% of its initial value when exposed to an intervention. The points are colored according to whether the intervention drives the internal BN of each cell to an attractor of the uncontrolled dynamics (blue) or not (orange); all interventions depicted result in activation of the apoptosis node. Attractor control, in all cases, is achieved by modulation of the S1P-SPHK1-PDGFR feedback loop (either directly or via the influence of Ceramide). }
    \label{fig:auc_vs_85_pulsed}
\end{figure}

In the second simulation scenario (\emph{global pulses}), shown in Figure \ref{fig:pulsed}, some interventions result in more immediate lethality but lower long-term effectiveness than others. To investigate this further, we compared the time required to drive the population to 85\% of its initial value against the integrated population level for each successful intervention (Figure \ref{fig:auc_vs_85_pulsed}). Data points are labeled by by whether they drive the system into a stable attractor of the uncontrolled dynamics (``Attractor Controls'') or not (``Target Controls''). These latter interventions induce apoptosis by creating a new attractor that does not exist in the uncontrolled dynamics. We see two distinct groups in Figure~\ref{fig:auc_vs_85_pulsed}, with one shifted along the integrated population axis relative to the other. The points with the lower integrated population for a given time to reach 85\% population correspond to ``slow and steady'' interventions that have a higher long-term lethality than might be expected from their initial performance.

All of these slow and steady interventions influence the SPHK1-S1P-PDGFR feedback loop either directly, or by setting Ceramide to the ON state, which inactivates S1P. In all cases, this influence is sufficient to inactivate the three nodes. The inactive state of this feedback loop is a stable motif, meaning it is self-stabilizing and none of the three nodes can be reactivated except by one of the other two. The stable motif analysis of pystablemotifs~\cite{rozumPystablemotifsPythonLibrary2022} indicates that triggering this stable motif is sufficient to guarantee eventual convergence into the apoptotic attractor. In other words, the cell is committed to apoptosis as soon as the stable motif is triggered, even if the apoptosis node has not yet been activated--no recovery is possible.

In contrast, the ``fast'' interventions that do not belong to this ``slow and steady'' group fail to ensure convergence into the apoptotic attractor of the uncontrolled dynamics. Instead, they merely ensure that the apoptosis node eventually activates when the intervention is applied for long enough. If the intervention is removed before the apoptosis node has been activated or the cell dies, it is possible for the cell to recover (whether or not this recovery occurs depends upon stochastic effects). 

The difference is most stark in Figure \ref{fig:auc_vs_85_pulsed}, but it also is apparent in Figures \ref{fig:pulsed} and \ref{fig:varying-bcs}, where it manifests as a crossing of population curves. Moreover, the Ceramide-S1P-SPHK1-PDGFR interventions exhibit monotonic population decline with minimal oscillation; cells continue to die long after the substrate has fallen below the lethal threshold. We conjecture that the oscillating behavior observed in the other interventions is due to recovery effects.

\subsection{Network distance correlation with experimental outputs}\label{sec:network_distance_results}

\begin{figure}
    \centering
    \includegraphics[width=0.5\linewidth]{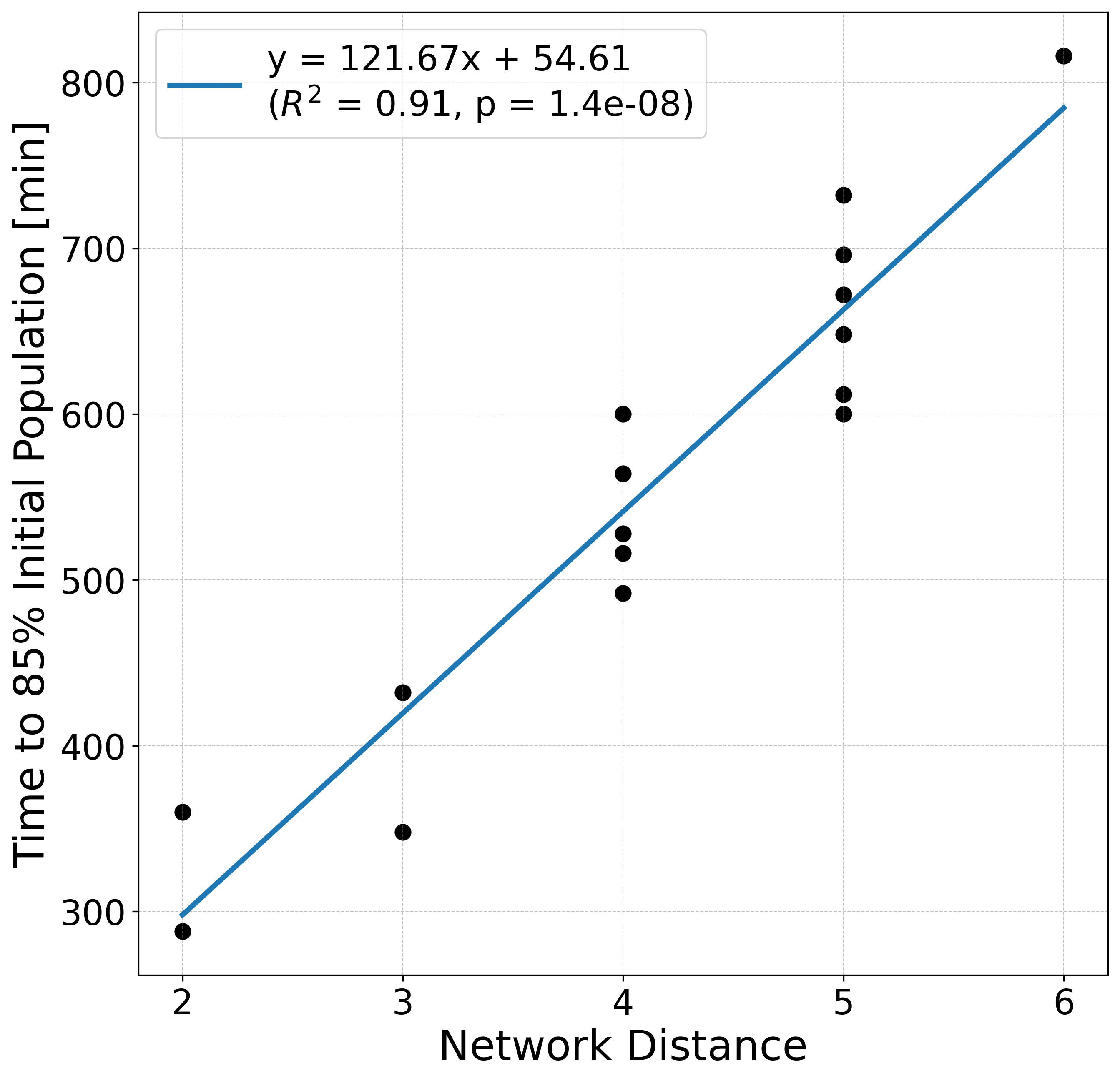}
    \caption{Dependence of initial lethality on network distance in the \emph{constant gradient} scenario. Each successful intervention is represented as a point. Network distance is computed as the length of the shortest path from the intervention target to the apoptosis node in the internal BN of each cell. The vertical position is the time (in minutes) between introduction of the apoptotic substrate and reduction of the population to 85\% of its initial value (interventions that never result in the population falling below this point are omitted). The least-squares regression is shown in blue.}
    \label{fig:auc_vs_dist}
\end{figure}

When the substrate concentration is held fixed, as in the first simulation scenario (\emph{constant gradient}, see Figure \ref{fig:constant-spatial}), the time it takes a successful intervention to reduce the population below some threshold depends on how quickly that intervention can propagate through the internal network of each cell. In Figure \ref{fig:auc_vs_dist}, we demonstrate that the network distance between the intervention and the apoptosis node is a dominant effect distinguishing how quickly the interventions take effect in the \emph{constant gradient} scenario. Specifically, we plot the time required to reduce the population to 85\% as a function of network distance. There are only two instances in which an intervention with a lower distance has a higher killing time than an intervention with a higher distance. 


\begin{figure}
    \centering
    \includegraphics[width=0.75\linewidth]{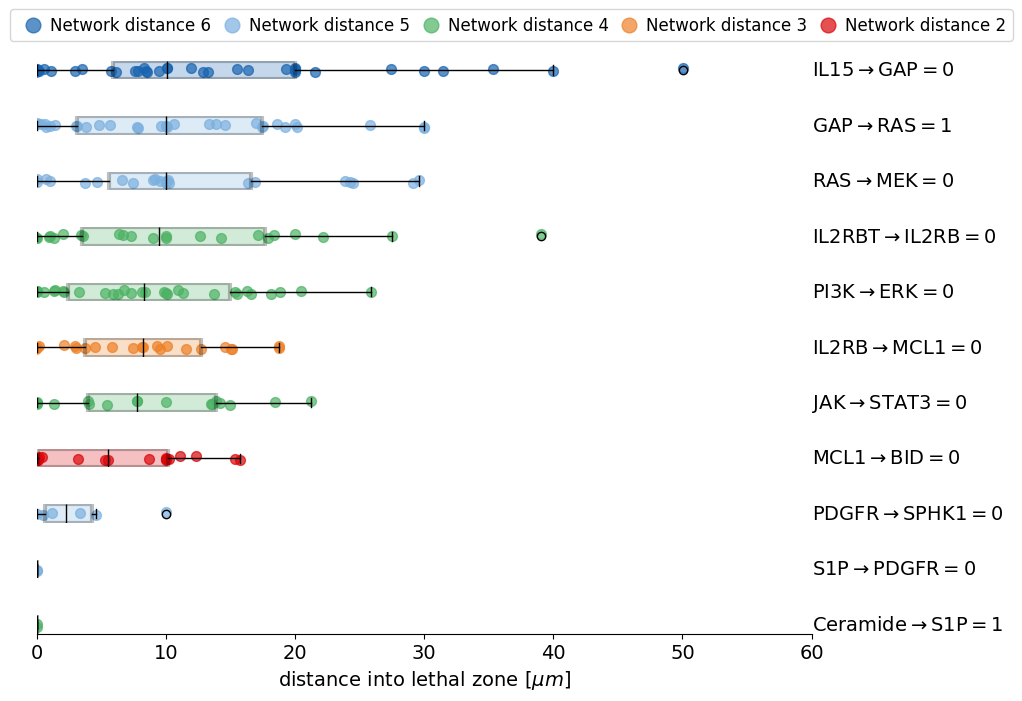}
    \caption{Infiltration of cells into the lethal zone of the \emph{time-varying boundary condition} scenario (slow decay, edgetic controls only). In the \emph{time-varying boundary condition} scenario, approximately half of the simulation domain is a lethal zone that is periodically exposed to sufficient concentration of apoptotic substrate to induce a reaction in cellular agents. Each row corresponds to a different substrate, colored by the distance between the apoptotic target and the apoptosis node in the internal BN of each node. The box plots of the spatial distribution of cells in the lethal zone are depicted, with individual points (cells) depicted. Vertical position is randomly chosen to aid in visualizing clusters of cells. Rows are ordered by the median (non-zero) penetration distance into the lethal zone.}
    \label{fig:lethal-dist}
\end{figure}

In the scenario with time-varying boundary condition, approximately half of the simulation region is potentially lethal. In general, the population extends slightly into this lethal zone and appears to establish a stable presence. However, the extent of this infiltration is not uniform across interventions, leading to separation in the final population levels observed in Figure~\ref{fig:varying-bcs}. We highlight this effect in Figure~\ref{fig:lethal-dist}, where we plot how far living cells have penetrated into the lethal zone for the 11 edgetic perturbations considered. The perturbations are ordered by the median penetration distance of cells in the lethal zone, and are colored by the distance in the regulatory network from the intervention target to the apoptosis node. The median penetration distance and network distance are in good agreement for target control interventions (top eight rows); only the orderings of the IL2RB$\rightarrow$MCL1 and JAK$\rightarrow$STAT3 interventions are in disagreement. In contrast, the attractor-control interventions that modulate the S1P-SPHK1-PDGFR feedback loop (bottom three rows) are substantially more lethal in this scenario than would be predicted by network distance alone. This is likely due to the reduced ability of cells affected by these interventions to recover, as discussed in section~\ref{sec:motif-effects}.

\section{Discussion}

Fully understanding the effects of interventions in living systems requires a multiscale approach. In a simple multicellular model of non-homeostatic leukemia survival, we have highlighted population-scale effects that emerge from nontrivial network effects at the sub-cellular level and developed a framework for exploring such effects in more detailed models.

Specifically, we evaluated the ability of a set of 28 controls inferred by three different methods to control dynamics to favor apoptosis. We find that some fail outright (predictions from the IBMFA method) whereas others produce time-dependent responses that are not easily understood in standard (implicit-time) Boolean simulation frameworks \cite{parkModelsCellProcesses2023, correiaCANAPythonPackage2018, albertBooleanNetworkSimulations2008, pauleveReconcilingQualitativeAbstract2020a}. We assess intervention efficacy through two measures - time to 85\% of initial population and total cell count. We find that the measures are generally correlated across scenarios. Furthermore, we explain aspects of observed population dynamics by categorization of the interventions into either attractor or target controls and by network distance between the target intervention node and the Apoptosis node. 

Crucially, the model studied here is relatively straightforward. For example, there is no direct cell-cell signaling (though cells do physically interact), uptake is modeled in a simple way, and Boolean time-scale parameters are not fine-tuned. Thus, we confidently propose that the behaviors we observe are general, and not dependent on specific features of the model considered here. In particular, qualitative differences in population response that arise from interventions targeting output nodes versus upstream feedback loops (stable motifs) does not rely on any fine-tuning of the model parameters or on special implementation details of the ABM component of the simulation. Thus, this simple example illustrates the importance of carefully considering internal agent dynamics in constructing ABMs. Moreover, more complex models may exhibit additional connections between internal and population dynamics; this possibility warrants further study.

This proof of concept integration and pipeline for setting and running computational experiments connecting network dynamics to multicellular simulations presents new opportunities to optimize experimental design prior to \textit{in vitro} and \textit{in vivo} experiments. To this end, future work will tune simulation parameters, such as the discrete-to-continuous transition functions, to data relevant to the system of study. Additional future work for this evaluation pipeline includes incorporation of a damage model to the cells, enabling, for example, the study of treatment resistance at the intracellular level. With interdisciplinary collaboration, this can hasten the development of deployable biomedical solutions along the translational spectrum \cite{leonard-dukeMultiscaleModelsLung2020, craigPracticalGuideGeneration2023, metzcarImprovingNeuroendocrineTumor2025}.

\section{Acknowledgements}

We thank Vincent Noël, Marco Ruscone, and Arnau Montagud for helpful discussions. This research was supported in part by Lilly Endowment, Inc., through its support for the Indiana University Pervasive Technology Institute.

\bibliography{johns_library, group_library}

\clearpage

\section{Supplementary Material}

\end{document}